\documentclass[10pt]{iopart}
\usepackage{graphicx}
\usepackage{array}
\usepackage{color}

\begin{document}

\title[First-principles calculations of the magnetic and electronic structures of MnP under pressure]{First-principles calculations of the magnetic  and electronic structures of MnP under pressure}

\author{Yuanji Xu$^{1,2}$, Min Liu$^{1,3}$, Ping Zheng$^1$, Xiangrong Chen$^3$, Jin-guang Cheng$^1$, Jianlin Luo$^{1,2,4}$, Wenhui Xie$^5$ and Yi-feng Yang$^{1,2,4,*}$}

\address{$^1$Beijing National Laboratory for Condensed Matter Physics and Institute of Physics, Chinese Academy of Sciences, Beijing 100190, China}
\address{$^2$School of Physical Sciences, University of Chinese Academy of Sciences, Beijing 100049, China}
\address{$^3$College of Physical Science and Technology, Sichuan University, Chengdu 610065, China}
\address{$^4$Collaborative Innovation Center of Quantum Matter, Beijing 100190, China}
\address{$^5$Department of Physics, Engineering Research Center for Nanophotonics and Advanced Instrument, East China Normal University, Shanghai 20062, China}
\ead{yifeng@iphy.ac.cn}

\begin{abstract}
Manganese monophosphide (MnP) shows complicated magnetic states varying with both temperature and pressure. We calculate the magnetic and electronic structures of MnP at different pressures using first-principles methods and obtain spiral ground states whose propagation vector changes from the $c$-axis at low pressure to the $b$-axis at high pressure. In between, we find a ferromagnetic state, as observed in the experimental phase diagram. The propagation vector of the spiral states is found to vary nonmonotonically with pressure, consistent with neutron measurements. Our results indicate that the complicated magnetic phase diagram originates from a delicate competition between neighboring exchange interactions between the Mn-ions. At all pressures, the electronic structures indicate the existence of quasi-one-dimensional charge carriers, which appear in the ferromagnetic state and become gapped in the spiral state, and anisotropic three-dimensional charge carriers. We argue that this two-fluid behavior originates from the special crystal structure of MnP and may be relevant for understanding the pairing mechanism of the superconductivity at the border of the high pressure spiral phase.
\end{abstract}


\vspace{2pc}

\submitto{\JPCM}

\maketitle
 
\ioptwocol

\section{Introduction}
\begin{figure}[t]
\includegraphics[width=0.5\textwidth]{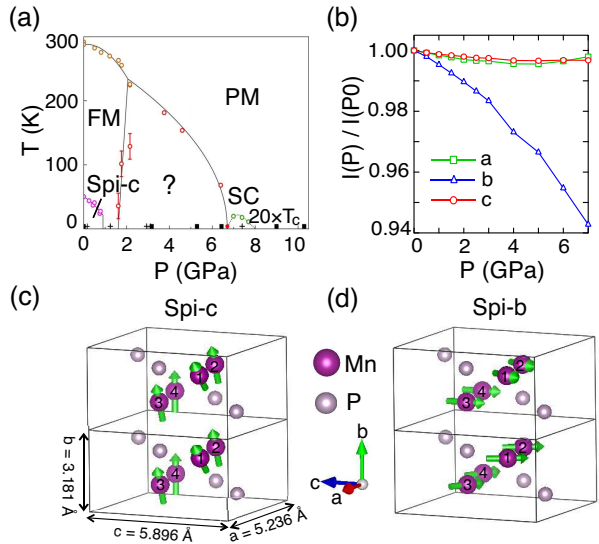}
\caption{(a) The magnetic phase diagram of MnP \cite{Matsuda2016,Wang2016}. The detailed spin structure of the high pressure state remains controversial. (b) Pressure dependence of the normalized lattice parameters determined from experiment, with $a$(P0)=5.2361 \AA{}, $b$(P0)=3.1807 \AA{} and $c$(P0)=5.8959 \AA{} at ambient pressure (P0) \cite{Wang2016}. (c) The spin structure of the low pressure Spi-c phase with the spins rotating in the $ab$ plane and propagating along the $c$-axis. (d) The spin structure of a candidate high pressure Spi-b phase with the spins rotating in the $ac$ plane and propagating along the $b$-axis.}
\label{figpd}
\end{figure}

Recently, MnP was found to display superconductivity with $T_c\approx 1\,$K at pressure around 7-8 GPa near the border of a long-range magnetic phase \cite{Cheng2015}. As the first Mn-based superconductor, it immediately raises the question concerning its pairing mechanism \cite{Rice2015}. The fact that superconductivity emerges at the border of a long-range magnetic order points to the possibility of magnetic glues due to spin fluctuations \cite{Scalapino2012} as has been proposed for the CrAs superconductivity \cite{Wu2014,Kotegawa2014}, but first-principles calculations in the framework of the weak-coupling BCS (Bardeen-Copper-Schrieffer) theory could also yield the correct transition temperature $T_c$ based solely on the electron-phonon coupling \cite{Chong2016}. The issue therefore remains controversial. Moreover, we still lack a good theoretical understanding of the associated magnetic and electronic structures, possibly due to the complicated phase diagram involving multiple magnetic orders varying with pressure and temperature, as shown in figure~\ref{figpd}(a) \cite{Matsuda2016,Wang2016}.

The magnetic orders in MnP have been a subject of many experimental and theoretical studies since 1960s \cite{Huber1964,Forsyth1966,Komatsubara1970}. At ambient pressure, it has an orthorhombic structure with $Pnma$ space group and the lattice constants $a = 5.236$, $b = 3.181$ and $c = 5.896$ \AA{} \cite{Wang2016}. Neutron diffraction experiment has provided very precise determination of its magnetic structure and revealed a paramagnetic (PM) to ferromagnetic (FM) phase transition at about 292 K and a ferromagnetic to spiral phase transition at about 47 K \cite{Yamazaki2014}. In the ferromagnetic phase, all spins align in parallel to the $b$-axis, whereas in the spiral phase, the spins rotate within the $ab$ plane with a propagation vector $\textbf{Q}=(0, 0, 0.117)$, indicating a periodicity of about nine lattice units along the $c$-axis (hereafter called Spi-c), as shown in figure~\ref{figpd}(c) \cite{Felcher1966}. Inelastic neutron scattering experiments on spin wave excitations \cite{Yano2013,Itoh2014} suggest that the spiral order may originate from the competition between different types of exchange interactions, as proposed in earlier theoretical work \cite{Takeuchi1967}. Under pressure, the spiral phase is first suppressed and replaced by a ferromagnetic ground state in a narrow pressure window around 1.2 GPa. Above 1.5 GPa, a new anti-ferromagnetic-like order appears and is then suppressed at very high pressure (7-8 GPa). Superconductivity emerges near the magnetic quantum critical point \cite{Cheng2015}. 

The magnetic structure of the high pressure phase has been extensively studied using neutron powder diffraction (NPD) \cite{Matsuda2016}, magnetic X-ray diffraction (XRD) \cite{Wang2016}, nuclear magnetic resonance (NMR) \cite{Fan2016}, and muon-spin rotation ($\mu$SR) \cite{Khasanov2016}, but the results remain controversial. While NPD and $\mu$SR indicate that the propagation vector changes from $c$-axis at ambient pressure to $b$-axis (Spi-b) at high pressure, XRD suggests that it remains along the $c$-axis but with a short periodicity. On the other hand, the NMR results imply a spiral structure at 2 GPa, but the exact magnetic structure cannot be resolved. The spin structure of the candidate high pressure Spi-b phase is shown in figure~\ref{figpd}(d) \cite{Matsuda2016}. The complicated magnetic phase diagram is probably related to the change of the lattice parameters with pressure. As shown in figure~\ref{figpd}(b), while both $a$ and $c$-axises change only slightly, the $b$-axis lattice parameter was found to decrease dramatically with pressure \cite{Wang2016,Selte1976}. However, the crystal symmetry remains the same without a structural phase transition. Previous numerical calculations have yielded  spiral phases at low and high pressures, but failed to produce the experimental \textbf{Q}-vector as well as the ferromagnetic phase at intermediate pressures \cite{Bonfa2016}. Moreover, the calculated energy differences between both spiral states and the ferromagnetic state are very small. More elaborate studies are needed in order to establish a systematic understanding of the variation of the magnetic orders in the experimental phase diagram.

The electronic structures of MnP have also been investigated recently at ambient pressure \cite{Zheng2016}. Detailed analysis of the optical spectra in comparison with first-principles calculations suggests the existence of two different types of charge carriers with distinct lifetimes. The short lifetime carriers originate from the $d_{y^2}$-orbital, exhibit quasi-one-dimensional character due to hybridization with the P $p$-orbitals, and are apt to order magnetically, whereas the long lifetime carriers consist of other Mn orbitals and are mainly responsible for the charge transport. It has been speculated that the interplay between these two types of carriers may be crucial if superconductivity emerges from the magnetic instability. However, it is not clear if this two-fluid property holds true at high pressures, although it is expected to be a property of the crystal structure which remains unchanged with pressure. A pressure-dependent investigation of the electronic structures is demanding.

In this work, we study the pressure evolution of the magnetic and electronic structures of MnP using first-principles density functional theory (DFT) with both the conventional collinear WIEN2k code \cite{Blaha} and the non-collinear WIENNCM code \cite{Laskowski2004}. We derive the exchange interactions from collinear calculations and predict correctly the spiral-ferromagnetic-spiral phase transitions with pressure. The results are further compared with non-collinear calculations, which yield all three magnetic phases as a function of pressure. The \textbf{Q}-vector is found to decrease with pressure for the Spi-c phase until it becomes zero (ferromagnetic) and then increase with pressure for the high pressure Spi-b phase, in good agreement with the overall trend observed in neutron experiment \cite{Matsuda2016}. We further show the coexistence of anisotropic three-dimensional (3D) Fermi surfaces and quasi-one-dimensional (1D) Fermi surfaces, supporting the existence of two types of charge carriers even at all pressures. The quasi-1D Fermi surfaces only exist in the ferromagnetic state and become gapped in the spiral states. Its interplay with the more itinerant 3D charge carriers may be the key to understand the electron pairing of the superconductivity, in resemblance of those in heavy fermion superconductors \cite{Yang2014}.

\section{Computational methods}
The electronic structure calculations were carried out with the WIEN2k \cite{Blaha} and WIENNCM \cite{Laskowski2004} packages using full-potential linearized augmented plane-wave and local orbital methods. We took the experimental lattice parameters under pressure as shown in figure~\ref{figpd}(b) but with relaxed internal coordinates \cite{Wang2016}. The Perdew-Burke-Ernzerhof generalized gradient approximation (GGA) was used for the exchange-correlation functional with 1500 k-point meshes for the whole Brillouin zone \cite{Perdew1996}. The Muffin-tin radii are set to 2.17 a.u. for Mn and 1.93  a.u. for P according to the high pressure structure. For non-collinear calculations using the WIENNCM code, the generalized Bloch wave function of the spiral spin structure takes the form:
\begin{eqnarray}
\psi_{k}\left(\mathbf{r}\right)=e^{i\mathbf{k}\cdot
\mathbf{r}}\left(\begin{array}{c} e^{\frac{-i\mathbf{Q}\cdot
\mathbf{r}}{2}}u_{k}^{\uparrow }\left(\mathbf{r}\right)\\
e^{\frac{i\mathbf{Q}\cdot \mathbf{r}}{2}}u_{k}^{\downarrow
}\left(\mathbf{r}\right)\end{array}\right),
\label{eq1}
\end{eqnarray}
which takes into account the periodicity of both the crystal and spin structures. The computational time is therefore greatly increased compared to the collinear calculations.

\section{The collinear calculations}
\begin{figure}
\includegraphics[width=0.5\textwidth]{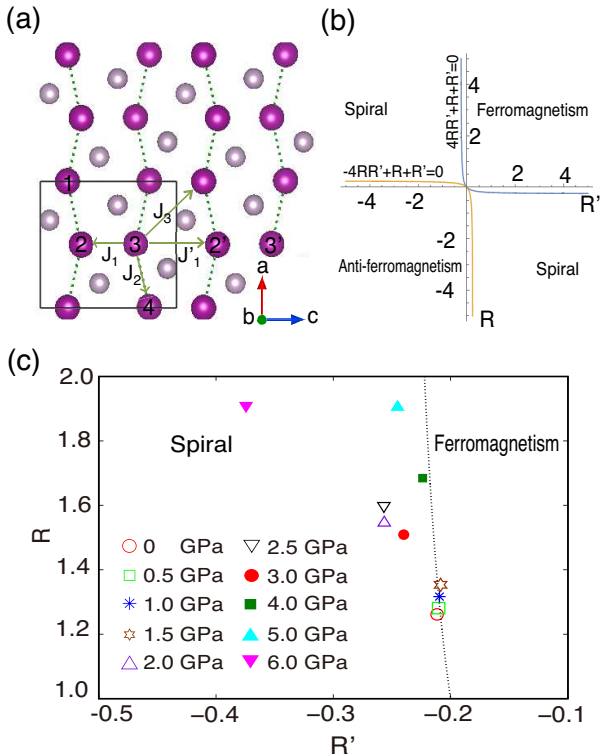}
\caption{(a) Illustration of the relevant exchange interactions with labeled Mn-ions. (b) The theoretical $R-R^\prime$ phase diagram for the magnetic ground state of the MnP-type structure, in which $R=J_1/J_2$ and $R^\prime=J^\prime_1/J_2$ \cite{Kallel1974,Bertaut1962}. The hyperbolic lines separate the ferromagnetic (or anti-ferromagnetic) and spiral phases. (c) An enlarged part of (b) showing the calculated values of $R$ and $R^\prime$ for MnP under pressure. The dashed line is given by the hyperbolic curve $4RR^\prime+R+R^\prime=0$ separating the ferromagnetic phase and the spiral phase.
}
\label{figsp}
\end{figure}

Spiral spin structure typically arises from magnetic frustration \cite{Kubler2000,Elliot1963,Moriya1960,Dzayloshinskii1958}. In MnP-type compounds, it involves three major exchange interactions between neighboring Mn-ions as shown in figure~\ref{figsp}(a) \cite{Kallel1974}, where $J_1$ and $J^\prime_1$ are the inter-chain exchange interactions and $J_2$ is the intra-chain exchange interaction along the Mn zigzag chain. The fourth-nearest-neighbor exchange interaction $J_{3}$ is found to be two orders of magnitude smaller in our calculations due to the relatively longer Mn-Mn distance and is therefore neglected in the discussions. The magnetic ground state is then determined by two dimensionless ratios, $R=J_1/J_2$ and $R^\prime=J^\prime_1/J_2$, which gives a theoretical phase diagram in figure~\ref{figsp}(b) \cite{Kallel1974,Bertaut1962}. In this simple model, the ferromagnetic (or anti-ferromagnetic) phase and the spiral phase are separated by two hyperbolic curves given by $4RR^\prime+R+R^\prime=0$ and $-4RR^\prime+R+R^\prime=0$.

To determine the property of the magnetic ground, we therefore need all three exchange interactions \cite{Yano2013,Itoh2014}. We consider four different magnetic structures as shown in table~\ref{table1} and use DFT to calculate their respective energy \cite{Gercsi2010}. The calculations involve a 1$\times$1$\times$2 supercell in order to obtain two inter-chain exchange interactions. Using the Heisenberg model $H=-\sum_{i,j}J_{ij}\textbf{S}_i\cdot\textbf{S}_j$, we have
\begin{eqnarray}
    E(FF)= -4J_1S^2 - 4J^\prime_1S^2 - 4J_2S^2 + E_0  \nonumber\\
    E(AF)= -4J_1S^2 - 4J^\prime_1S^2 + 4J_2S^2 + E_0  \nonumber\\
    E(FA)=  4J_1S^2 + 4J^\prime_1S^2 - 4J_2S^2 + E_0  \nonumber\\
    E(AC)= -4J_1S^2 + 4J^\prime_1S^2  + E_0
\end{eqnarray}
where $E_0$ is the referenced energy and $S$ is the obtained magnitude of the Mn-spins for each configuration.

\begin{table}%[!hbp]
\begin{tabular}{|p{1.5cm}<{\centering}|p{0.6cm}<{\centering}p{0.6cm}<{\centering}|p{0.6cm}<{\centering}p{0.6cm}<{\centering}|p{0.6cm}<{\centering}p{0.6cm}<{\centering}|}
\hline
Mn  atom & Mn1 & Mn2 & Mn3 & Mn4 & Mn2' & Mn3'  \\
\hline
FF & $\Uparrow $ & $\Uparrow $ & $\Uparrow $   & $\Uparrow $& $-$  &  $-$  \\
AF & $\Uparrow $ & $\Downarrow $ & $\Downarrow $   & $\Uparrow $& $-$   & $-$ \\
FA & $\Uparrow $ & $\Uparrow $ & $\Downarrow $   & $\Downarrow $& $-$   & $-$ \\
\hline
\hline
AC & $\Downarrow $ & $\Uparrow $ & $\Uparrow $   & $\Uparrow $& $\Downarrow $ & $\Downarrow $  \\
\hline
\end{tabular}
\caption{Four spin configurations used in the collinear calculations to determine the exchange interactions. The Mn-ions are labeled in figure~\ref{figpd}. }
\label{table1}
\end{table}

The resulting values of $J_1$, $J^\prime_1$ and $J_2$ are listed in table~\ref{table.table2} for different pressures. We see that $J_1$ and $J_2$ are both ferromagnetic, whereas $J^\prime_1$ is antiferromagnetic. The mean-field transition temperature for the ferromagnetic phase is then estimated to be $\sim$830$\,$K \cite{Anderson1963}, which is, as expected, 2-3 times higher than the experimental value of about 292$\,$K. In the literature, the origin of the exchange interactions has been ascribed to either the double exchange, the superexchange or the RKKY interactions \cite{Gribanov1983,Dobrzynski1989,Takase1979}. While we cannot provide a decisive answer to this issue, our DFT calculations do seem to suggest a correlation between the values of these exchanges interactions and their corresponding Mn-P-Mn bond angles. We find that for the ferromagnetic $J_1$ and $J_2$ the angles are between 73.4$^\circ$-74.0$^\circ$ and 69.6$^\circ$-70.8$^\circ$, respectively. Both are smaller than 90$^\circ$. While for the antiferromagnetic $J^\prime_1$, the Mn-P-Mn bond angle lies between 111.6$^\circ$-112.9$^\circ$ and is larger than 90$^\circ$. This seems to accord well with the Goodenough-Kanamori rule \cite{Goodenough1958,Kanamori1959}, which states a competition between the ferromagnetic double exchange mechanism and the antiferromagnetic superexchange interactions mediated by the P $p$-orbitals. The nonmonotonic pressure-dependence of $J_2$ seems to be correlated with the nonmonotonic variation of its Mn-P-Mn bond angle with increasing pressure. In fact, the local minimum of $J_2$ occurs at around 2.0-2.5 GPa where the corresponding Mn-P-Mn bond angle  takes its local maximum (69.8$^\circ$). On the other hand, $J_1$ increases as its Mn-P-Mn bond angle decreases with increasing pressure below 5 GPa.

The dimensionless ratios, $R$ and $R^\prime$, are then calculated and compared with the theoretical phase diagram in figure~\ref{figsp}(c). We find a spiral ground state below 1.0 GPa and above 2.0 GPa, and a ferromagnetic ground state between 1.0 and 2.0 GPa. Above 6 GPa, the magnetic moment of the Mn-ions of all states is found to be suppressed rapidly with pressure, suggesting the transition to a nonmagnetic state \cite{Gercsi2010}. The obtained ranges correspond well with the experimental phase diagram shown in figure~\ref{figpd}(a) \cite{Wang2016} and seem to be correlated with the nonmonotonic variation of $J^\prime_1$ with pressure. This is a very delicate change. Both the low pressure spiral state and the intermediate ferromagnetic state are close to the phase boundary. The ferromagnetic state becomes the ground state when the ratio $|R^\prime|=-J^\prime_1/J_2$ takes its minimum and crosses slightly over the phase boundary given by $4RR^\prime+R+R^\prime=0$. This suggests that both states can be very sensitive to external perturbation \cite{Komatsubara1970,Yamazaki2014}. As a matter of fact, $\mu$SR experiment has observed the coexistence of the ferromagnetic and spiral states at intermediate pressures \cite{Khasanov2016}. On the other hand, the high pressure spiral phase seems stable and locates away from the phase boundary in the $R-R^\prime$ phase diagram, consistent with the large pressure range of the Spi-b phase \cite{Matsuda2016,Khasanov2016}. 

We conclude that the complicated magnetic phase diagram of MnP originates from a delicate competition of the magnetic exchange interactions between neighboring chains. However, we would also like to point out that the simple Heisenberg-type localized spin model is specially designed for the study of the spiral magnetic structure. While it may be valid for describing the basic features of the spiral state, some important factors are obviously missing, including non-Heisenberg-like terms such as the Dzyaloshinsky-Moriya (DM) interaction  \cite{Yamazaki2014} and the itinerant part of the Mn $d$-electrons \cite{Zheng2016}. The interplay of these terms may become important at a more delicate level and cause some interesting physics such as a topological Hall effect \cite{Shiomi2012}. One should therefore be cautious when using the simplified spin model to interpret complicated experimental data. A better model that contains both the localized and itinerant behavior of the Mn $d$-electrons will be needed in pursuit of a fully satisfactory solution of the MnP physics.

\begin{table}%[!hbp]
\begin{tabular}{|p{1.3cm}<{\centering}|p{1.5cm}<{\centering}|p{1.2cm}<{\centering}|p{1.2cm}<{\centering}|p{1.2cm}<{\centering}|}
\hline
P (GPa) & Order & $J_1$ & $J^\prime_1$ & $J_2$ \\
\hline
\hline
0.0 & Spiral &  83.664  & -14.022  &  66.311 \\
0.5 & Spiral &  84.855  & -13.929  &  66.249 \\
1.0 &   FM    &  86.620  & -13.783  &  65.790 \\
1.5 &   FM    &  88.290  & -13.606  &  65.245 \\
2.0 & Spiral &  92.099  & -15.269  &  59.515 \\
2.5 & Spiral &  94.045  & -15.128  &  58.899 \\
3.0 & Spiral &  96.250  & -15.306  &  63.791 \\
4.0 & Spiral & 102.568 & -13.628  &  60.899 \\
5.0 & Spiral & 109.868 & -14.139  &  57.675 \\
6.0 & Spiral &   85.068 & -16.684  &  44.559 \\
\hline
\end{tabular}
\caption{The calculated exchange interactions and the corresponding magnetic state for different pressures. The unit of all exchange interactions is meV.}
\label{table.table2}
\end{table} 

\section{The non-collinear calculations}
\begin{figure}[t]
\includegraphics[width=0.5\textwidth]{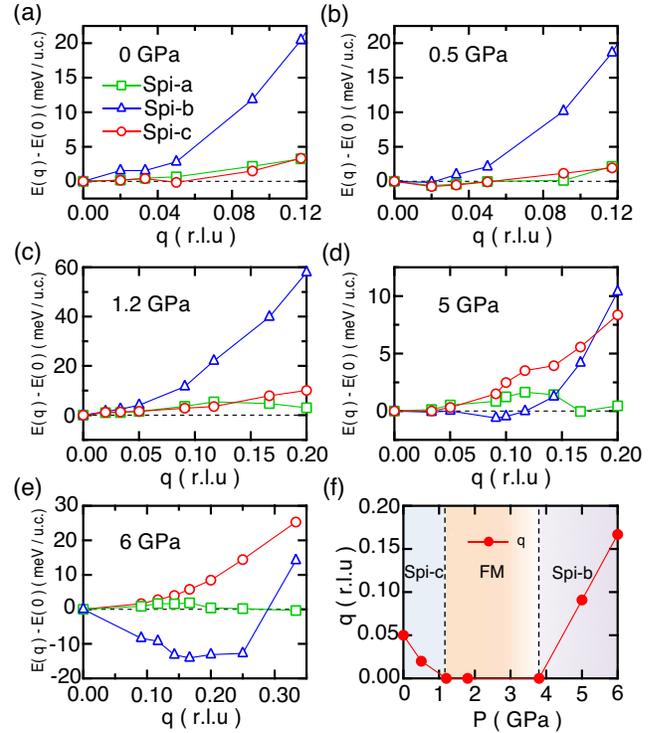}
\caption{(a-e) The energy difference between three spiral states and the ferromagnetic state ($q=0$) as a function of the propagation vector $q=|\textbf{Q}|$ at 0 GPa, 0.5 GPa, 1.2 GPa, 5.0 GPa and 6.0 GPa. E(0) is the energy of the ferromagnetic state calculated using the WIENNCM code. (f) The calculated pressure-dependence of $q$ for the ground state.
}
\label{figeq}
\end{figure}

The simplified model in the collinear calculations may not apply in the high pressure phase. To determine the detailed structure of the spiral phases with pressure, we performed non-collinear calculations using the WIENNCM code. At ambient pressure, we find that the spin rotation between Mn1 and Mn2 (or between Mn3 and Mn4) is about $21^{\circ}$ and that between Mn2 and Mn3 is about $2^{\circ}$, both in good agreement with experiment \cite{Forsyth1966}. Our pressure-dependent results are plotted in figure~\ref{figeq}, where we compare the energies as a function of $q=|\textbf{Q}|$ for three spiral structures with different propagation vectors: Spi-a with $\textbf{Q}=(q,0,0)$, Spi-b with $\textbf{Q}=(0,q,0)$, and Spi-c with $\textbf{Q}=(0,0,q)$. We find that the energy of the Spi-c state increases systematically with increasing pressure, whereas that of the Spi-b state decreases with increasing pressure, possibly owing to the rapidly reducing lattice parameter along the $b$-axis \cite{Wang2016}. Hence below 1.2 GPa, the Spi-c state has the lowest energy, whereas for pressure above 4 GPa, the Spi-b state has the lowest energy. At intermediate pressures, $q$ reduces to zero and we find a ferromagnetic ground state. 

Figure~\ref{figeq}(f) plots the pressure dependence of $q$ for the ground states. We see that $q$ decreases with increasing pressure in the Spi-c phase, remains zero in a finite pressure range, and then increases with pressure in the Spi-b phase. In neutron experiment, $\textbf{Q}$ is found to be (0, 0, 0.117) at ambient pressure, (0, 0 ,0) at 1.2 GPa, (0, 0.091, 0) at 1.8 GPa and (0, 0.141, 0) at  3.8 GPa \cite{Matsuda2016}. While the overall trend agrees well with our theoretical prediction, there seems to be a systematic mismatch in the pressure range, which may be attributed to the numerical discrepancy of the non-collinear calculations.  We also note that the high pressure Spi-b state has a much lower energy compared to the ferromagnetic state, in contrast to the small energy difference (only the order of 1 meV) between the Spi-c state and the ferromagnetic state at ambient pressure. 

As shown in our collinear calculations, the largest inter- and intra-chain couplings, $J_1$ and $J_2$, are both ferromagnetic and much higher in magnitude than other exchange couplings. They connect all the Mn-spins through zigzag paths and intend to form a ferromagnetic network covering the whole lattice. Therefore, the spiral magnetic structure can only result from frustrations introduced by  higher-order antiferromagnetic exchange couplings such as $J_1^\prime$. The evolution of the $q$-dependent energy profile at 5 and 6 GPa reflects the effect of increasing frustrations with pressure.

We briefly discuss the possible role of the spin-orbital coupling and the DM interaction in determining the magnetic structures of MnP. Recent neutron scattering experiments at ambient pressure have shown that the DM interaction may indeed play a role, causing a small tilt of the Mn spins from the $ab$-plane towards the $c$-axis \cite{Yamazaki2014} and yielding an unconventional Hall effect of possibly topological nature \cite{Shiomi2012}. However, this is a very delicate issue and goes beyond our simple calculations for the basic spiral structures. We have examined the results by including the spin-orbital coupling in our DFT calculations and obtained qualitatively similar pressure-dependence in the ${\bf Q}$-vector. The energy differences between $E(q)$ and $E(-q)$ for all spiral states are found to be only the order of 0.5 meV. While this might be the expected magnitude for the DM interaction, it may also arise from numerical errors as we do not find a simple $q$-dependence expected for the DM energy. All together, our results suggest that the spin-orbital coupling will not change the basic spiral structure of MnP. However, more accurate numerical calculations are needed in order to quantitatively understand its effect on the detailed magnetic structures as suggested by the neutron scattering data.

\section{The electronic structures}
\begin{figure}
\includegraphics[width=0.5\textwidth]{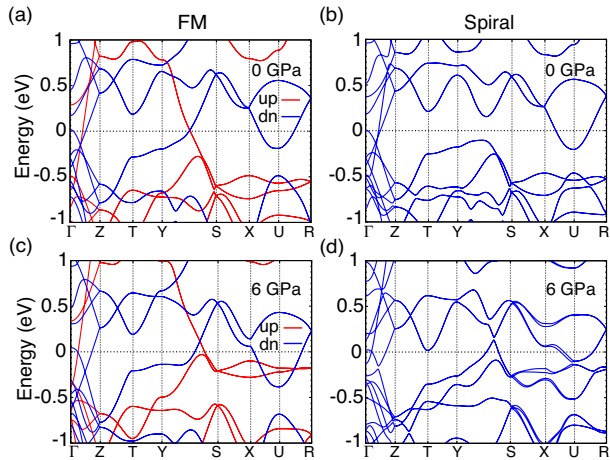}
\caption{The band structures for the Spi-c ($q=0.117$) and ferromagnetic states at 0 GPa \cite{Zheng2016} and the Spi-b ($q=0.167$) and ferromagnetic states at 6 GPa. The high symmetry points in the horizontal axis are indicated below in figure~\ref{figfs}(f). }
\label{figbd}
\end{figure}

\begin{figure}
\includegraphics[width=0.5\textwidth]{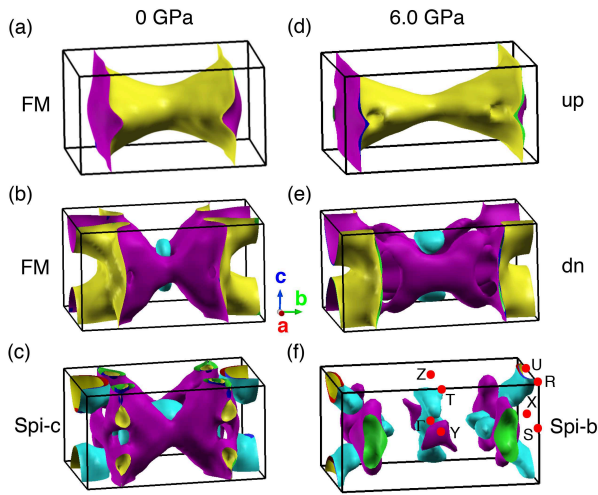}
\caption{The Fermi surfaces for the Spi-c ($q=0.117$) and ferromagnetic states at 0 GPa \cite{Zheng2016} and the Spi-b ($q=0.167$) and ferromagnetic states at 6 GPa. The points in (f) indicate the high symmetry points used in the band structure calculations.}
\label{figfs}
\end{figure}

The band structures and Fermi surfaces are calculated for both the spiral and ferromagnetic spin structures and compared in figures~\ref{figbd} and \ref{figfs} for both 0 GPa and 6 GPa \cite{Zheng2016}. In the ferromagnetic case, the Fermi surfaces consist of both flat Fermi sheets and anisotropic 3D-cylindric Fermi surfaces. The flat Fermi sheets come from the Mn-$d_{y^2}$ orbital in hybridization with the P 3$p$-orbitals \cite{Zheng2016,Yanase1980,Grosvenor2005}. In the band structures, this corresponds to the two intersecting bands along the Y-S direction. The anisotropic 3D-cylindric Fermi surfaces come from other Mn 3$d$-orbitals. As discussed previously \cite{Zheng2016}, these special topology of the Fermi surfaces give rise to two different types of charge carriers which coexist in the ferromagnetic state. In the spiral state, the quasi-1D Fermi surfaces are gapped, giving rise to several hole pockets scattered in the Brillouin zone, and the associated charge carriers contribute a major portion to the ordered moments. The gap opening in the spiral phase originates from the hybridization between the spin up and spin down channels caused by the magnetic scattering or the folding of the Brillouin zone associated by the \textbf{Q}-vector. No qualitative change is seen at 6 GPa except that the gap locates slightly above the Fermi energy, which indicates that the quasi-1D charge carriers are only partially gapped in the Spi-b phase.

As proposed in previous theoretical studies \cite{Lizarraga2004}, the special band crossing along the Y-S direction turns out to be particularly susceptible to non-collinear instabilities. This suggests a close connection between the electronic band structures and the spiral magnetic ground state, indicating that the Mn-$d_{y^2}$ orbitals are largely responsible for the spiral instability. For small $q=|\textbf{Q}|$, the hybridization gap between the spin up and spin down channels is determined roughly by \cite{Lizarraga2004}
\begin{eqnarray}
\Delta=\sqrt{(\overline{\textbf{v}}_\textbf{k} \cdot \textbf{Q})^{2}+4V^2},
\end{eqnarray}
where $\overline{\textbf{v}}_\textbf{k}$ is the average velocity of the two spin channels at the wave vector $\textbf{k}$ in the Brillouin zone and $V$ is the hybridization strength given by the off-diagonal elements of the Hamiltonian between the two components of the generalized spinor state in equation~(\ref{eq1}). The exact form of $V$ may depend on $\textbf{k}$ and $\textbf{Q}$. For the quasi-1D band along the Y-S line, we have $\overline{\textbf{v}}_\textbf{k}\cdot\textbf{Q}\approx 0$ in the Spi-c phase, hence the hybridization gap is roughly given by $2V$. The hybridization strength can therefore be estimated to be $V \sim 0.14\,$eV, based on the gap opening at the crossing point as shown in figure~\ref{figbd}(b). This is the same order of magnitude as the exchange interactions. On the other hand, for the Spi-b phase at 6 GPa, while the gap opening is small at the crossing point slightly above the Fermi energy, there is an overall band shift along the Z-T-Y line if we compare the band structures for the ferromagnetic state in figure~\ref{figbd}(c) and the Spi-b state in figure~\ref{figbd}(d). This should also be understood to arise from the hybridization effect. The overall magnitude of the band shift is about 0.2 eV, similar to that estimated for $V$ at 0 GPa. 

\begin{figure}
\includegraphics[width=0.5\textwidth]{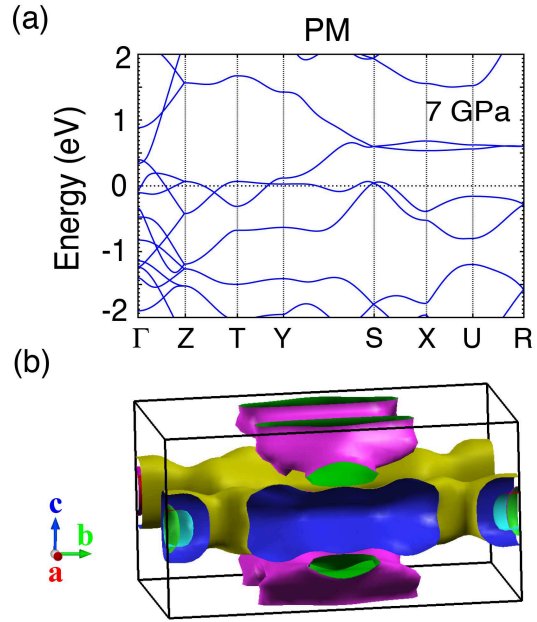}
\caption{The band structures and Fermi surfaces for the paramagnetic state at 7 GPa.}
\label{figpm}
\end{figure}

To see the change of carriers at higher pressures, we also calculate the band structures and Fermi surfaces for the paramagnetic state at 7 GPa and the results are plotted in figure~\ref{figpm}. It is seen that the quasi-1D bands along the Y-S direction are both shifted away from the Fermi energy so that the quasi-1D carriers will also not contribute to the charge transport as is the case in the spiral states. Neither will it contribute directly to the superconducting condensation. We speculate that the localized $d_{y^2}$-electrons will produce magnetic quantum critical fluctuations responsible for the superconducting pairing of other more itinerant 3D-like carriers since they provide a major contribution to the magnetic orders at low pressures. This might resemble the paramagnetic state in the localized regime of the two-fluid model in heavy fermion systems.

We would like to further comment on the origin of the quasi-1D character of the Mn $d_{y^2}$-orbital. Our magnetic calculations show that Mn-ions have a magnetic moment of 1.4 $\mu{_B}$ with a d4$\uparrow$2$\downarrow$ configuration, indicating a Mn$^{1+}$P$^{1-}$ valence state rather than Mn$^{3+}$P$^{3-}$ \cite{Tremel1986}. The former is in agreement with recent X-ray photoelectron spectroscopy measurements \cite{Grosvenor2005}. In such a case, the P-ions should be treated as zigzag P-P clustering chains containing [P$_2$]$^{2-}$ units with strong P-P interaction, instead of isolated P$^{3+}$ anions. We find that it is the P-P anti-bonding states that hybridize with the Mn-Mn bonding states along the $b$-axis and help to establish the quasi-1D dispersion of the Mn 3$d_{y^2}$-bands around the Fermi energy.

\section{Conclusion}
We have investigated the magnetic ground states using both collinear and non-collinear DFT calculations. We obtain both the spiral phases and the ferromagnetic phase as observed in the neutron experiment and predict correctly the pressure-dependence of the propagation vector. Our results indicate that the complicated magnetic phase diagram of MnP may be explained from the delicate competition between neighboring exchange interactions. The resulting electronic structures show characteristic quasi-1D Fermi sheets that become partially gapped in the spiral phase. This is a stable feature of the MnP-type structure and remains robust at high pressures. It suggests a two-fluid scenario which may provide an electronic basis for understanding the pairing mechanism of superconductivity at the border of non-collinear magnetism.

\ack
Y.X. thanks R. Laskowski for providing the non-collinear WIENNCM code. This work was supported by the National Natural Science Foundation of China (Nos. 11522435, 11574377), the State Key Development Program for Basic Research of China (2015CB921303, 2014CB921500), and the Strategic Priority Research Program (B) of the Chinese Academy of Sciences (XDB07020200, XDB07020100). Y.Y. was supported by the Youth Innovation Promotion Association CAS. X.C. was supported by the Science Challenge Project (Grant No. JCKY2016212A501) and the NSAF (Grant No. U1430117).

\end{document}